\documentclass[%
 reprint, 
 amsmath,amssymb,
 aps, prl,
]{revtex4-2}

\usepackage{graphicx}
\usepackage{dcolumn}
\usepackage{bm}

\usepackage[normalem]{ulem}
\usepackage{xcolor}

\begin{document}

\preprint{APS/123-QED}

\title{\textbf{
A source of heralded atom-photon entanglement for quantum networking
}}%

\author{Gianvito Chiarella$^{1}$}
\author{Tobias Frank$^{1}$}%
\author{Leart Zuka$^{1}$}%
\author{Pau Farrera$^{1,2}$}%
\email{Corresponding author: pau.farrera@mpq.mpg.de}
\author{Gerhard Rempe$^{1}$}%
\affiliation{$^{1}$Max-Planck-Institut für Quantenoptik, Hans-Kopfermann-Straße 1, 85748 Garching, Germany}
\affiliation{$^{2}$Munich Center for Quantum Science and Technology (MCQST), Schellingstr. 4, D-80799 München}

\date{\today}

\begin{abstract}
\noindent Communication in quantum networks suffers notoriously from photon loss. Resulting errors can be mitigated with a suitable measurement herald at the receiving node. However, waiting for a herald and communicating the measurement result back to the sender in a repeat-until-success strategy makes the protocol slow and prone to errors from false heralds such as detector dark counts. Here we implement an entanglement herald at the sending node by employing a cascaded two-photon emission of a single atom into two optical fiber cavities: The polarization of one photon is entangled with the spin of the atom, and the second photon heralds entanglement generation. We show that heralding improves the atom-photon entanglement in-fiber efficiency and fidelity to $68(3)\%$ and $87(2)\%$, respectively. We highlight the potential of our source for noise-limited long-distance quantum communication by extending the range for constant fidelity or, alternatively, increasing the fidelity for a given distance.
\end{abstract}

\maketitle


\noindent One of the biggest current challenges in quantum-information experiments and envisioned applications is reducing quantum-state errors \cite{Bennett2000,Preskill2018}. State infidelities and inefficiencies can be small in closed systems with isolated stationary qubits such as those employed in quantum computation \cite{Clark2021}. Quantum communication, however, is based on flying qubits and open systems that are assembled from individual devices, all of which suffer from such errors \cite{Gisin2007}. These errors are hard to eliminate but can be tracked and discarded by heralding the presence of the wanted quantum state. For example, a herald can distinguish successful from unsuccessful communication attempts and thereby make long-distance communication protocols more reliable \cite{Minar2012, Boone2015, Niemietz2021}. More fundamental, heralding is a key resource for measuring over long distances a detection-loophole-free violation of a Bell inequality \cite{Hensen2015}, something that is required for device-independent quantum key distribution \cite{Gisin2010, Cabello2012, Zhang2022}. Remarkably, and despite plenty of work on heralded photonic quantum states \cite{Bouwmeester1997, Barz2010, Kocsis2013, Hamel2014, Maring2017, Cao2024, Chen2024} and heralded quantum memories \cite{Kalb2015, Brekenfeld2020}, only little attention has been paid to the heralded generation of atom-photon entanglement \cite{Uphoff2016}. The latter, however, is a fundamental resource for quantum-information networks \cite{Kimble2008, Wehner2018} with entanglement established between remote atoms \cite{Moehring2007, Hofmann2012, Daiss2021, Knaut2024}.

While a typical herald signal is generated at the photon-receiving node, heralding the generation of atom-photon entanglement at the sending node is important for several reasons: First, the sender can immediately react upon errors and thereby avoid communication time in the network. Second, attenuation over the distance is avoided so that the herald signal can be obtained with a high signal-to-noise ratio. Third, arguably most important, the herald signal provides information about the time when the entanglement is generated. Communicating this information to the receiver allows one to know with high accuracy the arrival time of the entangled photon, with the advantage that faulty counts coming from random detector noise can be gated away. Heralding atom-photon entanglement directly at the source, therefore, has the potential to extend the communication range in quantum networks.

Here, we report on the heralded generation of entanglement between the polarization degree of freedom of a photon and the spin degree of freedom of a single atom. The entangled state is generated during a cascaded two-photon emission process of the atom \cite{Freedman1972, Aspect1981,Srivathsan2013} into two crossed optical cavities that collect photons efficiently \cite{Chiarella2024}. The photon emitted on the upper transition carries the polarization qubit that is entangled with the atomic spin, and the photon emitted in the lower transition is used as a herald for a successful entanglement generation process. The polarization eigenmode splitting \cite{Uphoff2015} of the herald cavity is essential for selectively steering the atom along one out of many possible decay pathways, thereby improving the fidelity of the entangled state. We show quantitatively how the herald signal allows tracking and reducing inefficiency and infidelity errors present during the entanglement generation process. We also observe that the herald photon carries information about the emission-time jitter present in the entangled photon, and we use this information to reduce photon qubit measurement noise. Finally, we show the potential of our source for noise-limited long-distance quantum networks.

\begin{figure}
\includegraphics[width=1\linewidth]{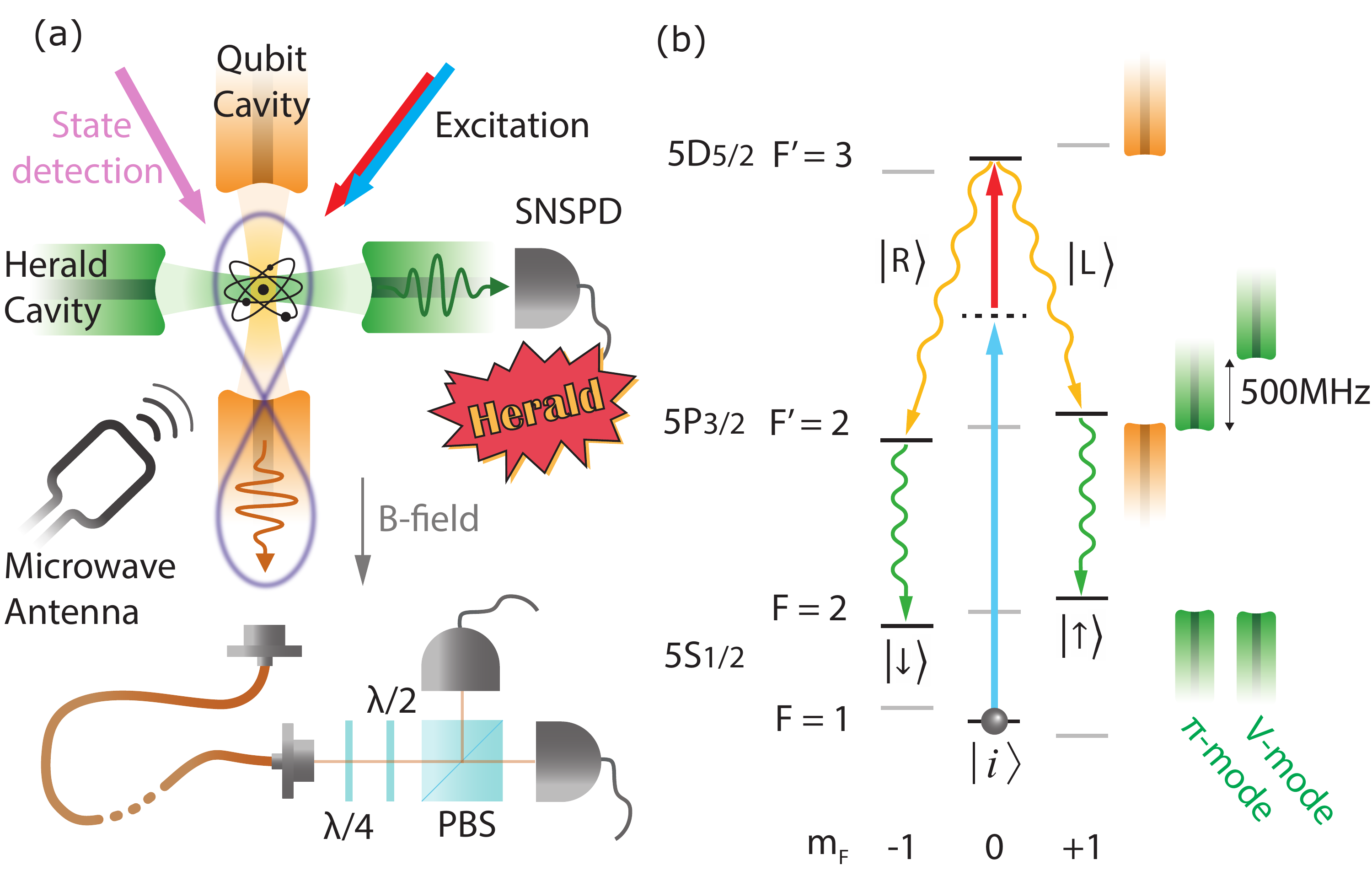}
\caption{\label{fig:setup}(a) A single atom coupled to two crossed fiber cavities emits a photon entangled with the atom and another photon that heralds the entangled state generation. Photon polarization and atomic qubit measurements are subsequently performed. (b) After optically pumping the atom into state $\left|5^2S_{1/2}, F=1,m_F=0\right>$, it is optically excited to state $\left|5^2D_{5/2}, F=3, m_F=0\right>$. A polarization qubit is generated in the qubit cavity, and the cavity-enhanced emission of a $\pi$ polarized photon in the herald cavity brings the atom into a superposition of two well-defined spin states.}
\end{figure}

Our experiment starts by optically cooling and trapping individual $^{87}\mathrm{Rb}$ atoms at the center of two crossed optical fiber cavities \cite{Brekenfeld2020}. The ``qubit cavity" couples the upper transition $\left|5^2D_{5/2}, F=3\right> \rightarrow \left|5^2P_{3/2}, F=2\right>$ and the ``herald cavity" couples the lower transition $\left|5^2P_{3/2}, F=2\right> \rightarrow \left|5^2S_{1/2}, F=2\right>$ (see Fig.~\ref{fig:setup}). After optically pumping the atom into state $\left| i \right> = \left|5^2S_{1/2}, F=1,m_F=0\right>$, two laser pulses induce a two-photon excitation to state $\left|5^2D_{5/2}, F=3,m_F=0\right>$, and the atom subsequently decays, emitting single photons into both cavities \cite{Chiarella2024}. Both cavities have one of the mirrors with a significantly higher transmission than the other one, such that photons exit the cavities dominantly through this mirror. During the experiment, we introduce a magnetic field pointing along the qubit cavity, such that the quantization axis is defined in that direction, and single photons in a superposition of right circular $\left|R\right>$ and left circular $\left|L\right>$ polarizations are emitted into this cavity. The mirrors of the herald cavity are elliptically curved such that the cavity is birefringent with nondegenerate polarization eigenmodes \cite{Uphoff2015}. The direction of the oscillating electric field of one polarization eigenmode (the $\pi$ mode) is oriented along the magnetic field so that this mode enhances the emission of $\pi$-polarized photons. This mode is resonant to the lower transition while the other herald cavity eigenmode (the V mode) is 500\,MHz blue detuned from this transition. In this situation, the two cavities enhance the emission of a single-photon polarization qubit in the qubit cavity that is entangled with the atomic spin and correlated with a $\pi$-polarized single photon (herald photon) emitted in the herald cavity. The entangled state can be expressed as
\begin{equation}
\left|\Psi\right\rangle = \frac{1}{\sqrt{2}}\left(\left|R,\downarrow\right\rangle+\left|L,\uparrow\right\rangle\right)
\label{eq:bellstate}
\end{equation}
where $\left|\uparrow\right\rangle=\left|5^2S_{1/2}, F=2,m_F=+1\right>$ and $\left|\downarrow\right\rangle=\left|5^2S_{1/2}, F=2,m_F=-1\right>$ denote the states of the atom.

\begin{figure}[hb]
\includegraphics[width=1\linewidth]{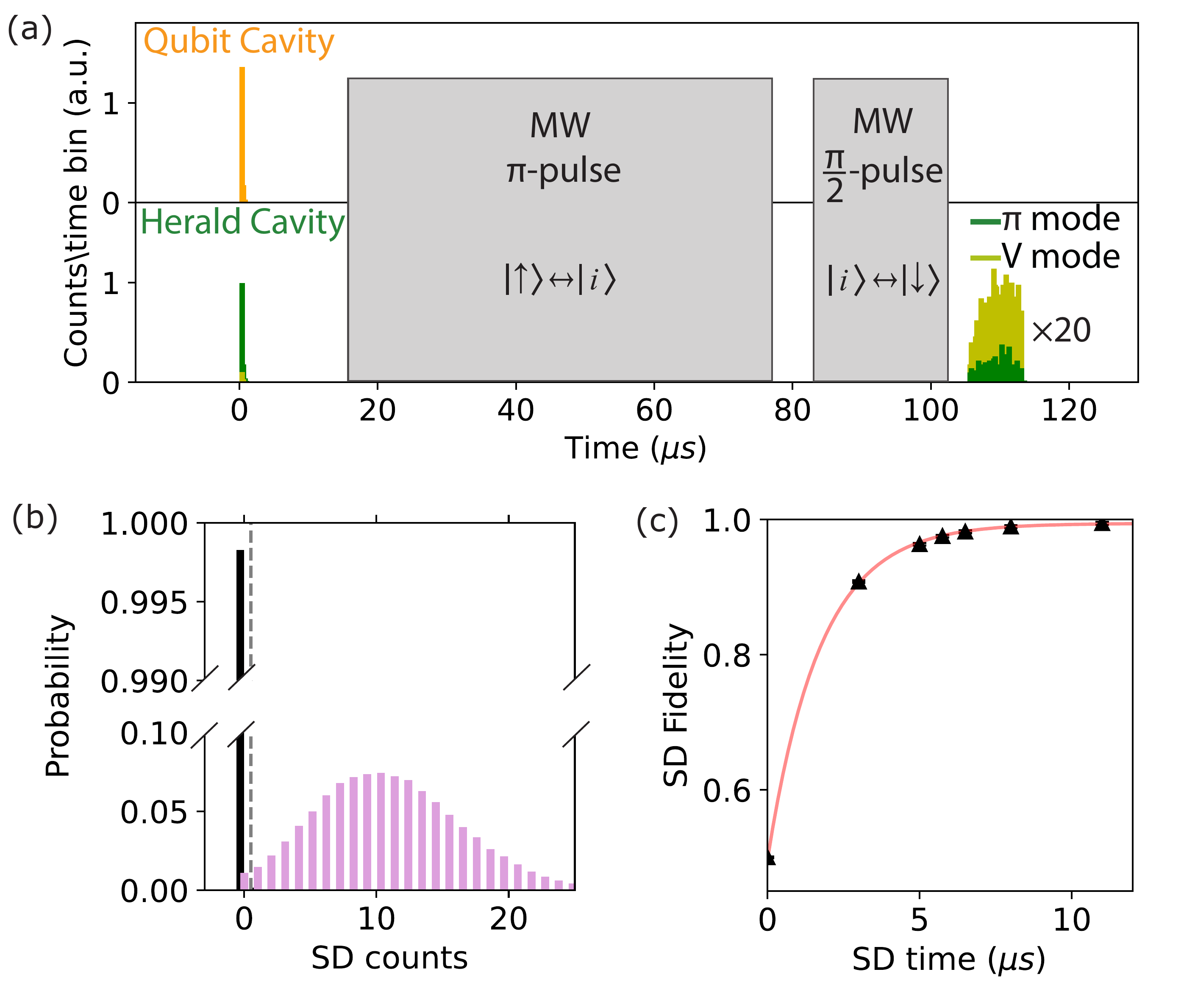}
\caption{\label{fig:readout} (a) Time histogram showing the photons detected in the qubit (upper plot) and herald (lower plot) cavities during the entanglement generation and measurement periods. The fluorescence counts are magnified by a factor 20. The grey area represents the microwave (MW) pulses used for the atomic qubit measurement. (b) Photon number histogram measured during the cavity-enhanced fluorescence atomic state detection when the atom is in state $\left|5^2S_{1/2}, F=1\right>$ (black bars) and in state $\left|5^2S_{1/2}, F=2\right>$ (pink bars). (c) Fidelity of the atomic state readout as a function of the measurement duration.}
\end{figure}

\begin{figure*}
\includegraphics[width=1\linewidth]{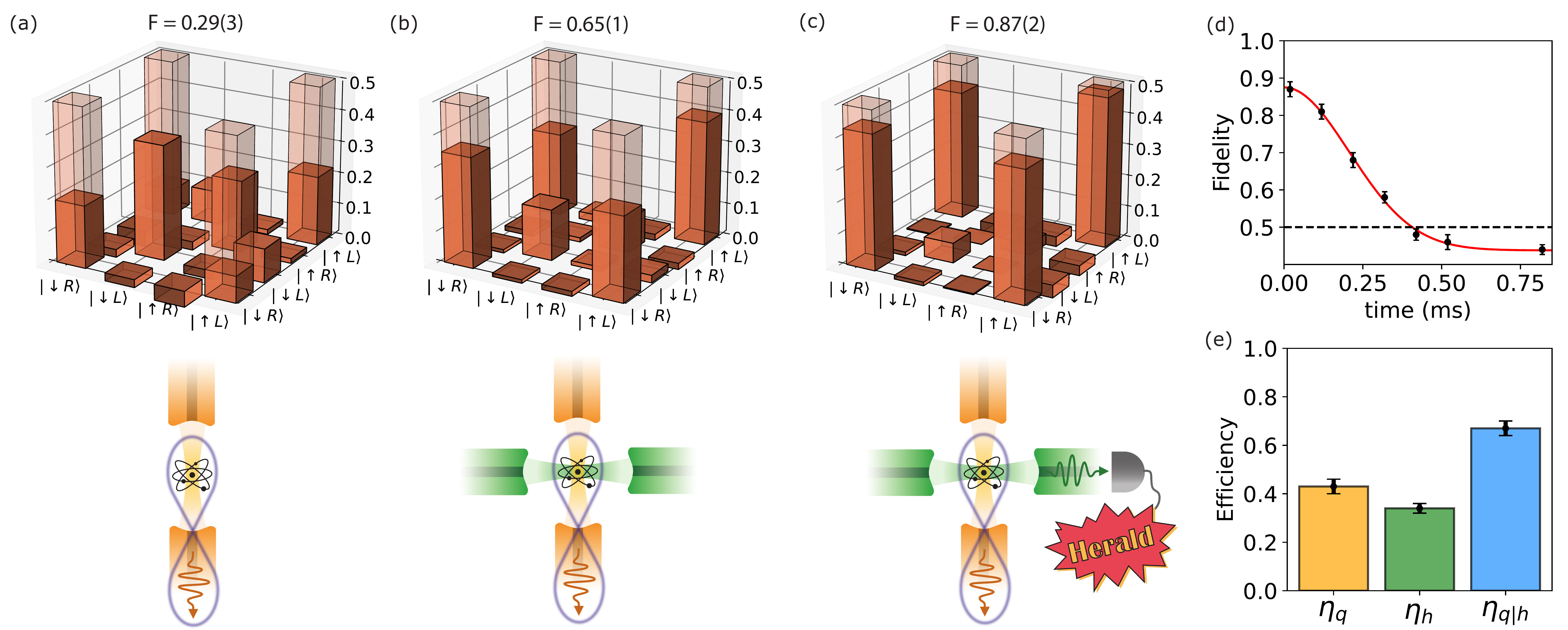}
\caption{\label{fig:tomography} Real part of the atom-photon density matrix corresponding to the situation where the herald cavity is not resonant to the lower transition (a), the herald cavity is resonant to the lower transition (b), the herald cavity is resonant to the lower transition and only events with a herald photon detection are considered (c). The shaded areas represent the ideal state in Eq.~\ref{eq:bellstate}. (d) Entangled state fidelity compared to an ideal Bell state as a function of the atomic qubit measurement-delay time. The dashed line represents the fidelity lower bound for an entangled state.(e) In-fiber efficiency to generate a qubit photon ($\eta_{q}$), a herald photon ($\eta_{h}$) and a heralded qubit photon ($\eta_{q|h}$).}
\end{figure*}

In order to analyze the entangled state, different qubit tomography techniques are employed. The single photon polarization qubit is analyzed with a combination of polarization wave plates, a polarizing beam-splitter, and single-photon detectors, allowing to perform polarization qubit measurements in an arbitrary basis. To perform atomic qubit measurements in an arbitrary basis, we use a sequence that employs microwave (MW) pulses and fluorescence state detection. Firstly, the atomic population in $\left|\uparrow\right>$ is transferred to $\left|i\right>$ with a microwave (MW) $\pi$ pulse  (see Fig.~\ref{fig:readout}a). Subsequently, a MW rotation is applied on the $\left|i\right>-\left|\downarrow\right>$ transition, in which the phase and amplitude of the MW pulse determine the basis of the qubit measurement. The measurement finishes by performing cavity-enhanced fluorescence atomic state detection by sending a laser 16\,MHz red-detuned to the $\left|5^2S_{1/2}, F=2\right> \leftrightarrow \left|5^2P_{3/2}, F=3, m_F = 3\right>$ cycling transition for 7.5 $\mu$s. This allows to distinguish if the atom is in state $\left|5^2S_{1/2}, F=1\right>$ or $\left|5^2S_{1/2}, F=2\right>$.  The fluorescence photons emitted during this process are enhanced by the two polarization eigenmodes of the herald cavity (the $\pi$ and V modes), which are detuned from the fluorescence laser by -255\,MHz and +245\,MHz, respectively. 

The complete atomic qubit readout sequence, including microwave rotations, delay intervals, and fluorescence-based state detection, is executed within 115 $\mu$s following the generation of the atom-photon entangled state. We perform $\pi$ MW rotations on the $\left|i\right> \leftrightarrow \left|\uparrow\right>$ transition with a fidelity of $\mathcal{F}=0.95(1)$ and $\pi$/2 rotations in the $\left|i\right> \leftrightarrow \left|\downarrow\right>$ transition with a fidelity of $\mathcal{F}=0.96(1)$ (these values include optical pumping and state detection errors). The cavity-enhanced fluorescence atomic state readout allows us to distinguish the presence of the atom in each of the two ground states by detecting either zero or up to about 20 photons in the herald cavity detector (see Fig.~\ref{fig:readout}b). The fidelity of the atomic state readout depends on its duration as it can be observed in Fig.~\ref{fig:readout}c. Extending the state detection duration allows the cavity to collect more fluorescence photons, thereby enhancing the fidelity of state readout. However, the cyclic absorption and emission of fluorescence photons heats the atoms, reducing their lifetime in the trap. In our experiment, we choose an atomic state detection duration of 7.5$\mu$s, corresponding to an average fidelity of $\mathcal{F}=0.990(2)$. 

Since the magnetic field splits the energy of atomic levels $\left|\uparrow\right\rangle$ and $\left|\downarrow\right\rangle$, the atomic state experiences a Larmor precession. After the measurement of the photonic qubit projects the atomic state, it will precess with a frequency $\nu_L=|E_{\uparrow}-E_{\downarrow}|/h=200\,\textrm{kHz}$ \cite{Wilk2007,Rosenfeld2008}. We can observe this precession in the oscillation of the atom-photon correlation parameter shown in the Supplementary Material.

In order to characterize the atom-photon entangled state and the impact of the herald photon, Fig.~\ref{fig:tomography}  shows the density matrix of the two-qubit state for three different situations. In (a), the herald cavity is not resonant to the lower transition, and all the events with a qubit photon detection are considered. In (b), the herald cavity is resonant to the lower transition, and all the events with a qubit photon detection are considered. In (c) the herald cavity is resonant with the lower transition, too, but only the events with a photon detection in both the qubit and the herald cavities are considered. We observe how the fidelity of the atom-photon entangled state compared to an ideal Bell state increases from $\mathcal{F}=0.29(3)$ in (a) to $\mathcal{F}=0.65(1)$ in (b), and to $\mathcal{F}=0.87(2)$ in (c), well above the fidelity bound for entangled states ($\mathcal{F}>0.5$) \cite{Sackett2000}. This increase is due to the enhancement of $\pi$ atomic decay on the lower transition, a process which is induced by the $\pi$ mode of the herald cavity, and which can be heralded by the detection of a $\pi$ polarized herald photon. The obtained heralded entangled state fidelity is limited by the decoherence of the atomic state during the atomic qubit measurement time. 

We also characterize the decrease of the entangled state fidelity as a function of the atomic-qubit measurement-delay time after the herald-detection window. We observe a Gaussian decay of the fidelity with a decay time of 206(6)$\mu s$ (see Fig.~\ref{fig:tomography}d). For this measurement each data point is taken with a different atomic qubit measurement basis, which compensates for the atomic state precession. The decay is limited by fluctuations of the magnetic field at the atomic position. 

In addition to enhancing the entangled state fidelity, the heralding also enhances its efficiency. The reason is that the herald signal allows to discard events in which an imperfect excitation or the finite atom-cavity cooperativity leads to the atom not following the photon emission path described in Fig.~\ref{fig:setup}b.  Fig.~\ref{fig:tomography}d shows the in-fiber photon emission efficiency for the qubit photon ($\eta_{q}=43(3)\%$), the herald photon ($\eta_{h}=34(2)\%$) and the heralded qubit photon ($\eta_{q|h}=68(3)\%$). We can observe that the herald signal enhances the efficiency by a factor of 1.5. These values are currently limited by the cavity-fiber mode matching, the cavity out-coupling efficiency and the finite atom-cavity cooperativity \cite{Chiarella2024}.

\begin{figure}
\includegraphics[width=1\linewidth]{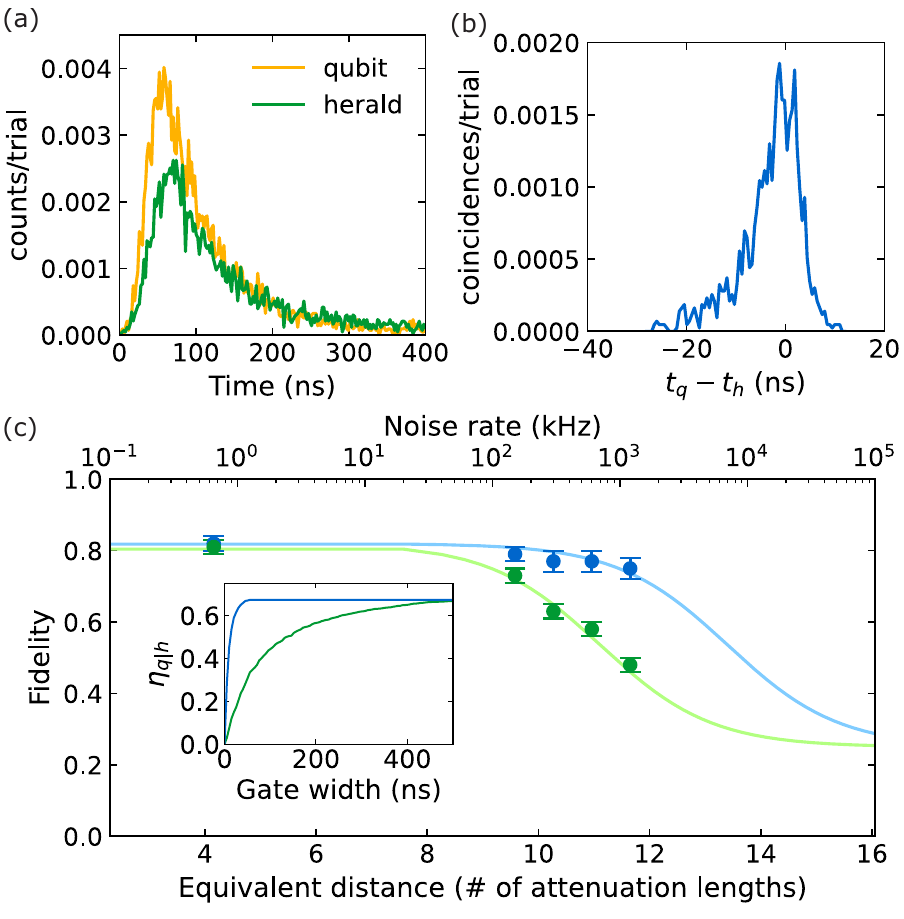}
\caption{\label{fig:comrate}(a) Qubit and herald photon detection time histogram. (b) Qubit-herald photon coincidence counts as a function of the photon detection time delay. (c) Entangled state fidelity as a function of the qubit measurement noise rate. For the green data points, the same photon qubit measurement gate is used for every heralded entanglement event (with gate duration $t_g=400ns$). The blue data points indicate the situation in which the herald photon detection time information is used in order to gate the qubit photon measurement (with $t_g=40ns$). The inset shows the heralded qubit photon efficiency as a function of the measurement gate duration ($t_g$). The blue data uses the herald photon arrival time in order to define the gate center and the green data uses a sequence trigger.}
\end{figure}

As mentioned above, the herald photon is temporally correlated with the qubit photon (see Figs.~\ref{fig:comrate}a,b). This is a consequence of the faster photon emission into the herald cavity compared to the qubit cavity, due to the stronger electric dipole moment on the $\left|5^2S_{1/2}, F=2\right> \leftrightarrow \left|5^2P_{3/2}, F=2\right>$ transition compared to the $\left|5^2P_{3/2}, F=2\right> \leftrightarrow \left|5^2D_{5/2}, F=3\right>$ transition \cite{Chiarella2024}. Thanks to this temporal correlation, the detection time of the herald photon tells information about the emission time of the qubit photon. One can then use this information in order to shorten the detector temporal gate used in the photon qubit measurement without a decrease in the heralded efficiency. As observed in the inset of Fig.~\ref{fig:comrate}c, shifting the gate according to the herald photon detection time (blue line) allows to have higher efficiency with shorter measurement gates, compared to the situation where the gate is not temporally shifted (green line). Reducing the measurement gate reduces the amount of detection noise that is mixed with the qubit measurement signal and improves the measured entangled-state fidelity in situations where noise is a limiting factor. 

Typically, measurement noise can become a limitation in situations where the signal propagates over a long distance and is significantly attenuated due to the transmission channel losses. Therefore in Fig.~\ref{fig:comrate}c we study the entangled state fidelity as a function of the amount of noise present in the photonic qubit measurement. We observe that temporally gating (with 40ns width) the photon qubit measurement with the information received from the detection time of the herald photon improves the state fidelity (blue data), compared to the situation where the full 400ns qubit photon gate is used (green data). The bottom horizontal axis shows the equivalent distance the photon could travel in an optical fiber in order to be measured with an equivalent signal-to-noise ratio (considering a qubit measurement noise rate of 10\,Hz).

In conclusion, we have shown the generation of a heralded atom-photon entangled state using a single atom coupled to two optical cavities. We have observed that heralding enhances the entangled-state generation efficiency and fidelity when compared to the unheralded case. Moreover, we have used the temporal correlation information between the qubit and herald photons in order to reduce the duration of the qubit measurement window. Making use of this feature, we have observed an improvement of the entangled state fidelity from $\mathcal{F}_{q|h}=48(2)\%$ to $\mathcal{F}_{q|h}=75(2)\%$ in a noise-limited photon-qubit measurement situation, which is relevant in long-distance quantum communication. 

We anticipate that our source will be useful in a quantum repeater architecture which we envision as consisting of heralded atom-photon entanglement sources, non-destructive qubit trackers \cite{Niemietz2021} and heralded quantum memories \cite{Brekenfeld2020}. In such an architecture, herald signals provided when generating, propagating and receiving entanglement will allow one to track and quickly react on source, channel and storage errors. All the mentioned components can be realized with single atoms trapped in crossed fiber cavities, highlighting the suitability of our system for next-generation entanglement-distribution schemes. Moreover, combining our heralding techniques with recently developed multiplexed entanglement sources based on registers of intra-cavity emitters \cite{Hartung2024,Krutyanskiy2024, Ruskuc2025} promises to reduce communication time and speed up repetition rates even further. Another promising feature of our scheme is the tunability of the qubit photon wavelength, determined by the chosen atomic transition. Specifically, bringing the qubit cavity to the $5^2P_{3/2}-4^2D_{3/2}$ transition of Rubidium would enable the generation of heralded atom-photon entangled states with the qubit photon wavelength in the low-loss telecom regime \cite{Uphoff2016}. This would allow us to build long-distance quantum networks without the need for quantum frequency conversion \cite{Radnaev2010, Albrecht2014, vanLeent2022, Liu2024}.

\begin{acknowledgments}
This work was supported by the Deutsche Forschungsgemeinschaft (German Research Foundation) under Germany’s Excellence Strategy – EXC-2111 – 390814868, and by the Bundesministerium für Bildung und Forschung (Federal Ministry of Education and Research) through projects QR.X (16KISQ019) and QR.N (16KIS2189). 
\end{acknowledgments}

\section{Appendix A: Atomic state manipulation}

The atomic qubit measurement has three steps. In the first step we transfer the full population in state $\left|\uparrow\right\rangle=\left|5^2S_{1/2}, F=2,m_F=1\right>$ to state $\left|i\right\rangle=\left|5^2S_{1/2}, F=1,m_F=0\right>$. In the second step atomic rotations on the transition between states $\left|i\right>$ and $\left|\downarrow\right\rangle=\left|5^2S_{1/2}, F=2,m_F=-1\right>$ are performed, which define the atomic qubit measurement basis. The final step is performing cavity-enhanced fluorescence atomic state detection in order to distinguish if the atom is in state $\left|5^2S_{1/2}, F=1\right>$ or $\left|5^2S_{1/2}, F=2\right>$. In order to characterize the overall performance of the measurement it is important to characterize the different steps independently. 

\begin{figure}[h]
\includegraphics[width=0.9\linewidth]{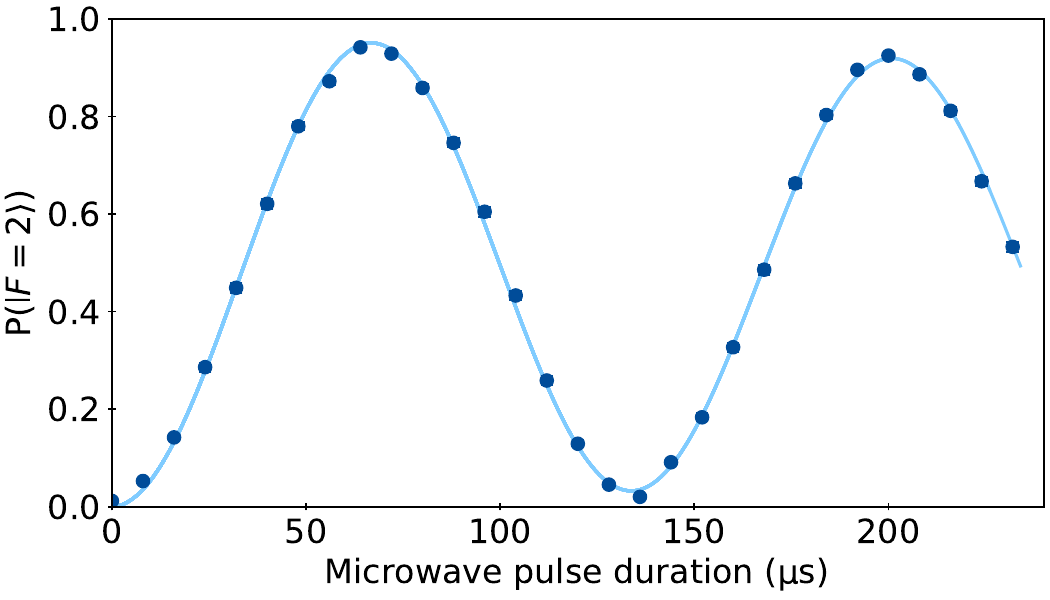}
\caption{\label{fig:rabism} A single atom initially in state $\left|i\right>$ interacts with microwave radiation that couples this state with state $\left| \uparrow\right>$.}
\end{figure}

Fig.~\ref{fig:rabism} shows atomic rotations on the $\left|i\right> \leftrightarrow \left|\uparrow\right>$ transition, obtaining a $\pi$-pulse fidelity of $\mathcal{F}=0.95(1)$. 

\begin{figure}[ht]
\includegraphics[width=0.9\linewidth]{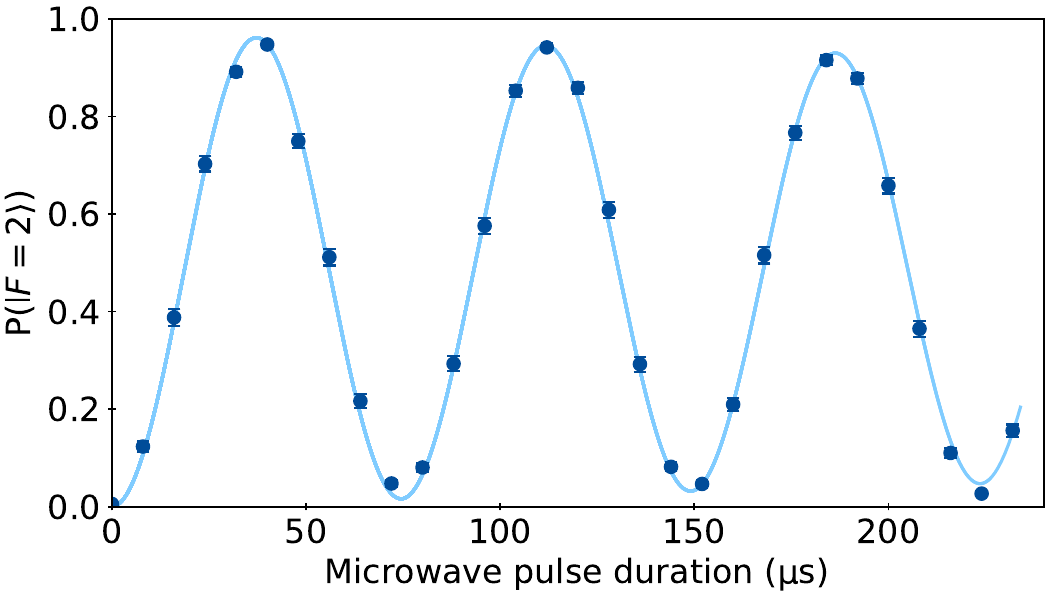}
\caption{\label{fig:rabisp} A single atom initially in state $\left|i\right>$ interacts with microwave radiation that couples this state with state $\left|\downarrow\right>$.}
\end{figure}

Fig.~\ref{fig:rabisp} shows atomic rotations on the $\left|i\right> \leftrightarrow \left|\downarrow\right>$ transition, obtaining a $\pi/2$ pulse fidelity of $\mathcal{F}=0.96(1)$.

The characterization of the final fluorescence atomic state detection step can be found in Fig.~2 in the main text.

\section{Appendix B: Atomic state precession}

\begin{figure}[ht]
\includegraphics[width=1\linewidth]{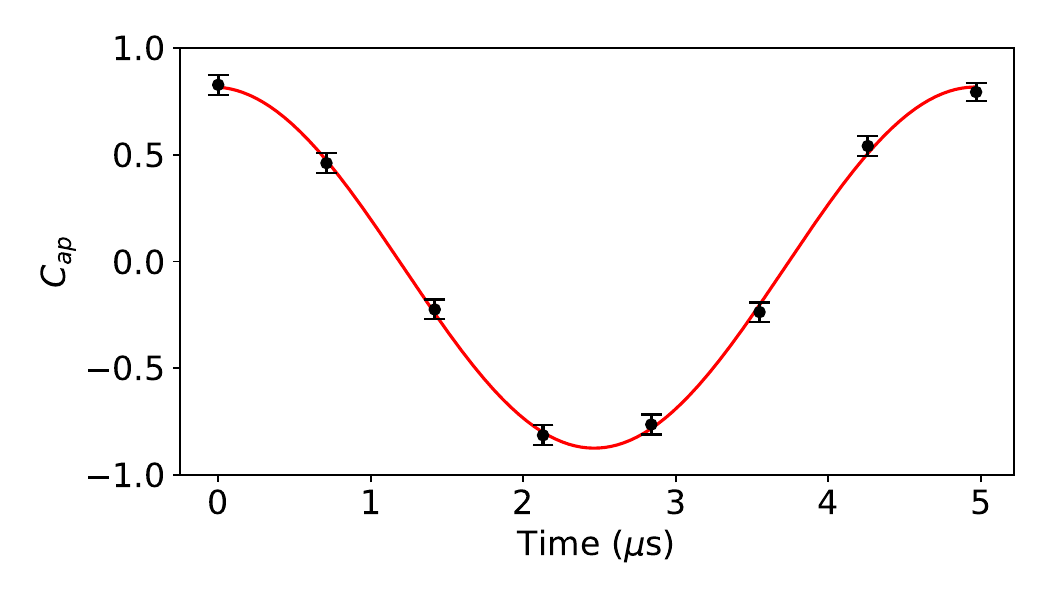}
\caption{\label{fig:atomicPrecession} Atom-photon state correlation parameter $C_{ap}$ as a function of the atomic qubit measurement time.}
\end{figure}

Since in our experiment we generate an homogeneous magnetic field at the position of the atom, the energy of atomic levels $\left|\uparrow\right\rangle=\left|5^2S_{1/2}, F=2,m_F=1\right>$ and $\left|\downarrow\right\rangle=\left|5^2S_{1/2}, F=2,m_F=-1\right>$ are split. In this situation, an atomic state that consists of a superposition of $\left|\uparrow\right\rangle$ and $\left|\downarrow\right\rangle$ will evolve over time in the form of a Larmor precession \cite{Wilk2007,Rosenfeld2008}. After the measurement of the photonic qubit projects the atomic state, it will precess with a frequency $\nu_L=|E_{\uparrow}-E_{\downarrow}|/h=200\,\textrm{kHz}$. We can observe this precession in the oscillation of the atom-photon state correlation parameter. This parameter is expressed as a function of the atom-photon state measurement probability P: 

\begin{equation}
C_{ap}=P(\uparrow_a,H_p)+P(\downarrow_a,V_p)-P(\uparrow_a,V_p)-P(\downarrow_a,H_p)
\label{eq:corr}
\end{equation}

where ($\uparrow_a$,$\downarrow_a$) denote the atomic spin state and ($H_p$,$V_p$) denote the horizontal and vertical polarization state of the photon. The evolution of this parameter as a function of the atomic qubit measurement time can be seen in Fig.~\ref{fig:atomicPrecession}. The period of the oscillations of $5 \mu s$ matches the expected value given by the splitting of the two atomic spin states.

\section{Appendix C: Entangled state fidelity in the presence of noise}
In order to simulate the data shown in Fig.~4c in the main text, we consider theoretically the situation in which an atom-photon entangled state is mixed with white noise in the photonic channel. This mixture can be considered in the atom-photon density matrix $\rho_{ap}$ as:

\begin{equation}
\rho_{ap}=p\rho^0_{ap}+(1-p)\rho^0_{a}\otimes \frac{\mathbb{I}_p}{2}
\label{eq:rho}
\end{equation}
where p is the fraction of events in which the measurement is performed on the original atom-photon state $\rho^0_{ap}$. $\rho^0_{a}$ is the remaining atomic state when the photonic state is traced out. When considering an atom-photon entangled Bell state mixed with white noise, $\rho^0_{a}=Tr_p(\rho^0_{ap})=\mathbb{I}_a/2$. Therefore we can rewrite the final atom-photon state as:

\begin{equation}
\rho_{ap}=p\rho^0_{ap}+(1-p)\frac{1}{4}\mathbb{I}_{a}\otimes\mathbb{I}_p
\label{eq:rho1}
\end{equation}

The fidelity of this state with respect to a Bell state is 

\begin{equation}
\mathcal{F}=p\mathcal{F}_0+\frac{1-p}{4}
\label{eq:fidelity}
\end{equation}

where $\mathcal{F}_0$ is the fidelity of $\rho^0_{ap}$ with respect to a Bell state. This equation is used to calculate the expected atom-photon entangled state fidelity in Fig.~4c in the main text.

\nocite{*}

\bibliography{publications-arxiv}

\end{document}